\documentclass[twocolumn,aps,prl,reprint,noshowpacs,letterpaper,floatfix,hypertext]{revtex4-2}
\usepackage{amsmath,amssymb}
\usepackage{txfonts}
\usepackage{bm}
\usepackage{textcomp}
\usepackage[pdftex]{graphicx,color}
\usepackage{braket}
\usepackage[]{hyperref}

\begin{document}
\title{Nonmagnetic Phase Transition in La$_3$TiSb$_5$ Observed in $^{121}$Sb Nuclear Quadrupole Resonance}

    \author{Masahiro Manago}
    \email{manago@riko.shimane-u.ac.jp}
    \author{Gaku Motoyama}
    \affiliation{Department of Applied Physics, Shimane University, Matsue 690-8504, Japan}
    \author{Shijo Nishigori}
    \affiliation{ICSR, Shimane University, Matsue 690-8504, Japan}
    \author{Kenji Fujiwara}
    \affiliation{Department of Applied Physics, Shimane University, Matsue 690-8504, Japan}
    \author{Hisatomo Harima}
    \affiliation{Department of Physics, Kobe University, Kobe 657-8501, Japan}

\begin{abstract}
    We measured the electrical resistivity and the $^{121}$Sb nuclear quadrupole resonance (NQR) spectra
    of La$_3$TiSb$_5$, which is a nonmagnetic reference system for a locally noncentrosymmetric
    antiferromagnet Ce$_3$TiSb$_5$, in order to determine its physical properties.
    The resistivity exhibited a hump anomaly at 120 K, suggesting a phase transition.
    The NQR spectra split below 110 K, whereas the nuclear spin-lattice relaxation exhibited no anomalies.
    These results can be explained by the incommensurate charge-density-wave (CDW) transition.
    This finding suggests the importance of quasi-one-dimensional characteristics and CDW instability
    in these systems.
\end{abstract}
\maketitle

Locally noncentrosymmetric systems have received considerable attention because various unusual phenomena
can be realized, for example, exotic superconductivity\cite{PhysRevB.93.224507,PhysRevB.98.224510}
and cross-correlation responses\cite{PhysRevB.90.024432,JPSJ.83.014703}.
Cross-correlation responses are expected to occur when odd-parity multipoles emerge
in conjunction with the magnetic transitions\cite{PhysRevB.90.024432,JPSJ.83.014703}.
The magnetoelectric effect, that is, current-induced magnetization, is an example of an unusual response
and is confirmed for two antiferromagnetic metals
UNi$_4$B\cite{JPSJ.87.033702} and Ce$_3$TiBi$_5$ \cite{JPSJ.89.033703,JPSCP.30.011189}.
The origin of the magnetoelectric effect in these systems remains under active discussion.
Understanding the magnetic structure is crucial to elucidate its mechanism.

\begin{figure}
    \centering
    \includegraphics{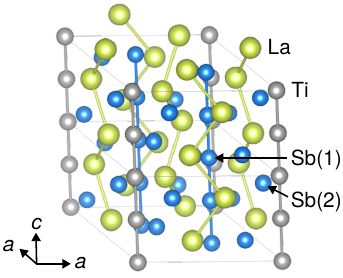}
    \caption{\label{fig:crystal}(Color online)
        Crystal structure of La$_3$TiSb$_5$ with a hexagonal $P6_3/mcm$ space group symmetry, drawn by VESTA\cite{JApplCryst.44.1272}.
        Two unit cells are stacked along the $c$ axis.
        The lattice constants are from literature\cite{Chem.Mater.7.2229}.
        Two nonequivalent Sb sites exist and are denoted as Sb(1) for $4d$ ($3.2$) position and Sb(2) for $6g$ ($m2m$).
        La and Sb sites lack the local inversion symmetry, whereas the Ti site is located on the inversion center.
    }
\end{figure}

Antiferromagnet Ce$_3$TiBi$_5$ belongs to the series of hexagonal RE$_3$TiX$_5$ systems
(RE: rare-earth element; X = Sb, Bi)
and these materials crystallize in a hexagonal $P6_3/mcm$ (No.~193, $D_{6h}^{3}$) space group symmetry\cite{Chem.Mater.7.2229,PhysicaB.536.142}.
Rare-earth atoms form a zigzag chain along the $c$ axis possessing no inversion symmetry.
Ce$_3$TiBi$_5$ exhibits a cycloid order with moments lying in the plane of the zigzag chain
below $T_{\textrm{N}}=5.0$ K\cite{PhysRevB.109.L140405}.
By contrast, Ce$_3$TiSb$_5$ exhibits an antiferromagnetic transition at $T_{\textrm{N}}=5.2$ K
with ordered moments along the $c$ axis with long-period modulation\cite{J.Phys.33.245801}.
Another remarkable feature of RE$_3$TiX$_5$ systems is the quasi-one-dimensional chain along the $c$ axis.
This is common in the nonmagnetic reference system La$_3$TiSb$_5$,
and Sb(1) and the Ti atoms are aligned along the $c$ axis, as shown in Fig.~\ref{fig:crystal}.
This could lead to the instability of the Fermi surface,
resulting in charge-density-wave (CDW) or spin-density-wave transitions.
Among the RE$_3$TiX$_5$ systems, La$_3$TiSb$_5$ exhibits an electrical resistivity anomaly
of unknown origin at approximately 120 K\cite{Chem.Mater.14.4867}.
Some phase transition is expected in La$_3$TiSb$_5$.

In this letter, we report the results of the electrical resistivity and $^{121}$Sb nuclear quadrupole resonance (NQR) measurements
on La$_3$TiSb$_5$ to reveal the anomalies at lower temperatures.
The resistivity exhibited a hump anomaly at 120 K.
The NQR spectrum began to split below the transition temperature of $\sim 110$ K.
This split was explained using a simple model that assumed spatially modulated electric field gradient (EFG)
because of the incommensurate CDW transition.
No clear anomaly was observed in the nuclear spin-lattice relaxation rate $1/T_1$.
This finding highlights the importance of the quasi-one-dimensional character of these systems.

The single-crystalline La$_3$TiSb$_5$ sample was synthesized using the Sn-flux method,
as in Ref.~\onlinecite{J.Phys.Cond.Matt.29.145601}.
La (99.9\%), Ti (99.5\%), Sb (99.99\%), and Sn (99.999\%) in molar ratios
of La:Ti:Sb:Sn = 2--3:1:5:20--50
were placed in an alumina crucible and sealed in a quartz tube.
The ingredients were heated to 1000\textdegree C, maintained at this temperature for 10 h,
and subsequently cooled to 500\textdegree C at $\sim -1.5$\textdegree C/h.
Needle-shaped crystals were obtained.
The excess Sn flux was removed by centrifugation.
The sample was cleaned with dilute hydrochloric acid to remove residual Sn flux.
Electrical resistivity measurements were performed on the single-crystal samples using
the standard four-terminal method.
The sample was crushed into a powder for NQR measurements to enhance the signal intensity.
In addition, the powdered sample was sealed in epoxy (Stycast 1266) in a rod shape to prevent oxidization.
The standard spin-echo method was used for NQR measurements.
The frequency-swept spectra were obtained by summing multiple Fourier-transformed spin-echo signals
at various frequencies.

\begin{figure}
    \centering
    \includegraphics{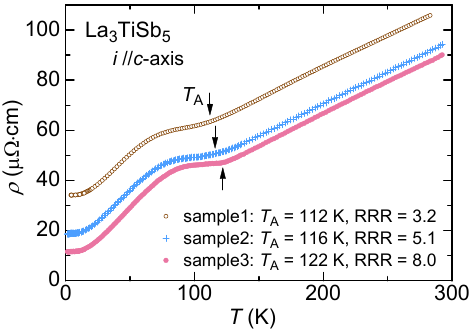}
    \caption{\label{fig:resistivity}(Color online)
        Temperature dependence on the electrical resistivity of single-crystalline La$_3$TiSb$_5$ samples.
        The current was applied along the $c$ axis.
        The hump anomaly is observed at approximately 120 K for the best sample.
        The vertical arrows indicate transition temperatures
        determined from the change in slope at the kink.
    }
\end{figure}

Figure \ref{fig:resistivity} shows the electrical resistivity of La$_3$TiSb$_5$ samples along the $c$ axis
in the range of 2--290 K.
A hump anomaly was observed at $\sim 120 $ K for the best-quality sample,
which agrees with a previous study\cite{Chem.Mater.14.4867}.
This suggests a reduction in the density of states (DOS) at the Fermi energy level.
The samples used in this study did not exhibit zero resistivity because of the superconductivity of the Sn flux below
4 K\cite{Chem.Mater.14.4867,J.Phys.Cond.Matt.29.145601}, although some samples exhibited a slight decrease at this temperature.
The sample quality varied to some extent, and the residual resistivity ratio (RRR) $\rho(\textrm{300 K})/\rho(\textrm{2 K})$
was between 3 and 8.
Furthermore, higher RRR samples exhibited higher and sharper anomalies.
Another sample from the same batch as the electrical resistivity was used for the NQR measurements.

\begin{table}
    \begin{center}
        \caption{\label{tab:nuq}
            Quadrupole frequency $\nu_{\textrm{Q}}$ and asymmetric parameter $\eta$
            from the band calculation and the $^{121}$Sb NQR experiment at the Sb(1) and Sb(2) sites in La$_3$TiSb$_5$.
        }
        \begin{tabular}{cccc}
            \hline\hline
                                                & $\nu_{\text{Q}}$ (MHz) & $\eta$ \\\hline
            $^{121}$Sb(1) (calc.)               & 66.7                   & 0      \\
            $^{121}$Sb(2) (calc.)               & 15.6                   & 0.4219 \\
            $^{121}$Sb(1) (NQR expr. at 300 K)  & 65.3                   & -      \\
            $^{121}$Sb(2) (NQR expr. at 5--7 K) & 15.9                   & 0.38   \\
            \hline\hline
        \end{tabular}
    \end{center}
\end{table}

\begin{figure}
    \centering
    \includegraphics{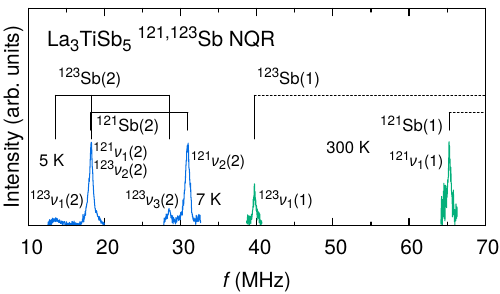}
    \caption{\label{fig:spectrum-assign}
        (Color online) $^{121}$Sb and $^{123}$Sb NQR spectrum of La$_3$TiSb$_5$.
        The spectrum of Sb(1) is at 300 K, and that of the Sb(2) is at 5 K or 7 K.
        The expected $^{123}$Sb(2) $\nu_2$ line coincides with the $^{121}$Sb(2) $\nu_1$ line at $\sim 18.3$ MHz.
        The signals of Sb(1) higher than 70 MHz were not measured in the present study.
        The spectra for the different frequency ranges were measured
        at different temperatures or with different NMR spectrometers, and the signal intensities are not comparable.
    }
\end{figure}

A nucleus with $I\ge 1$ possesses a nuclear quadrupole moment $Q$,
and the nuclear spin Hamiltonian without a magnetic field is
\begin{align}
    \mathcal{H}_{\textrm{Q}} = \frac{h\nu_{\textrm{Q}}}{6} \left\{
    \left[3I_z^2 - I(I+1)\right] + \frac{1}{2} \eta(I_{+}^2 + I_{-}^2)
    \right\},\label{eq:hamiltonian}
\end{align}
where $\nu_{\textrm{Q}}$ denotes the nuclear quadrupole frequency and $\eta$ denotes the asymmetric parameter.
These parameters are expressed by the local EFG $V_{ij}$ at the nuclear site:
$\nu_{\textrm{Q}}=3eQV_{zz}/[2I(2I-1)h]$ and $\eta = (V_{xx} - V_{yy})/V_{zz}$ when
the principal axes of the EFG are selected along the $x$, $y$, and $z$ axes, such that
$\lvert V_{zz} \rvert \ge \lvert V_{yy} \rvert \ge \lvert V_{xx} \rvert$.
Two isotopes of Sb exist: $^{121}$Sb ($I=5/2$) and $^{123}$Sb ($I=7/2$).
For $^{121}$Sb, two NQR transition lines
$\nu_1$ and $\nu_2$ are expected for each site.
The three-fold symmetry at the Sb(1) site ensures $\eta=0$, which leads to
simple resonant frequencies $\nu_1=\nu_{\textrm{Q}}$ and $\nu_2=2\nu_{\textrm{Q}}$
for the $\pm 1/2 \leftrightarrow \pm 3/2$ and $\pm 3/2 \leftrightarrow \pm 5/2$ transitions.
In contrast, $\eta$ is nonzero for the Sb(2) site because of the lower symmetry,
resulting in an $\eta$-dependent ratio between $\nu_1$ and $\nu_2$.
The $^{123}$Sb gives three lines per site, and the resonant frequencies are related to those of $^{121}$Sb
because these nuclei are located at the same crystal site.

Figure \ref{fig:spectrum-assign} shows the $^{121}$Sb and $^{123}$Sb NQR spectra at the selected temperatures.
The experimental $\nu_{\textrm{Q}}$ value for $^{121}$Sb was determined from the $\pm 1/2 \leftrightarrow \pm 3/2$
transition line for Sb(1) at $^{121}\nu_1(1)={}^{121}\nu_{\textrm{Q}}(1)$ ($\simeq 65.3$ MHz at 300 K).
Sb(2) parameters were evaluated based on the frequency of
$\nu_1(2)$ and $\nu_2(2)$ lines ($\sim 18$ and 31 MHz at 5--7 K, respectively) using numerical diagonalization
of the Hamiltonian in Eq.~\eqref{eq:hamiltonian}.
Table \ref{tab:nuq} lists the quadrupole parameters of Sb(1) and Sb(2) for the $^{121}$Sb nucleus
obtained through band calculation and the NQR experiment.
The band calculation result is in good agreement with the experimental values.
The ratio of $\nu_{\textrm{Q}}$ between $^{121}$Sb and $^{123}$Sb nuclei was
$^{121}\nu_{\textrm{Q}}(1)/^{123}\nu_{\textrm{Q}}(1)=1.64$ for Sb(1) site at 300 K, being
consistent with the expected value from the quadrupole moments\cite{AtData.111.1}:
$^{121}\{Q/[2I(2I-1)]\}/^{123}\{Q/[2I(2I-1)]\}=1.65$.
The signal intensity of $^{123}$Sb is weaker than that of $^{121}$Sb
due to the smaller gyromagnetic ratio and natural abundance, although it is unexpectedly weaker
compared to the typical relative intensity with that of $^{121}$Sb.
Hereafter, the NQR result for the $^{121}$Sb nucleus is shown.

\begin{figure}
    \centering
    \includegraphics{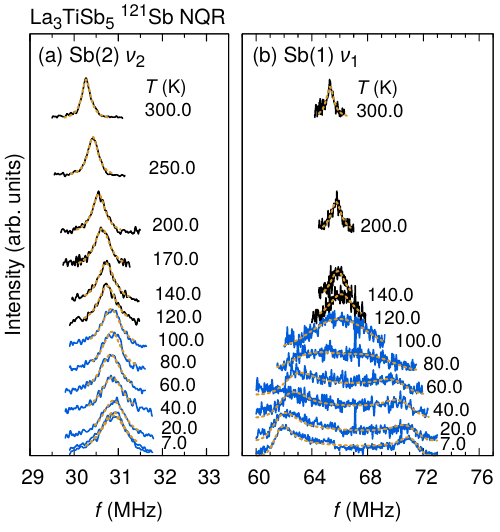}
    \caption{\label{fig:spectrum}(Color online)
        Temperature dependence of $^{121}$Sb NQR spectrum of (a)
        the $^{121}\nu_2(2)$ transition at the Sb(2) site and (b) $^{121}\nu_1(1)$ transition
        at the Sb(1) site in La$_3$TiSb$_5$.
        The spectra are shifted vertically.
        The dashed lines show fitting results with a Lorentzian function in the entire temperature range for (a)
        and at 120 K and above for (b);
        and a simulated spectral shape for the CDW state for 100 K and below for (b) (see text).
        The Lorentzian functions were obtained using the least-squares method, and the calculated spectra for the CDW state
        were selected manually to reproduce the experimental results.
    }
\end{figure}

Figure \ref{fig:spectrum} shows the NQR spectra at (a) Sb(2) $^{121}\nu_2(2)$ and
(b) Sb(1) $^{121}\nu_1(1)$ between 7--300 K.
The NQR spectra at $^{121}\nu_1(1)$ broadened below 100 K and split into two peaks at lower temperatures.
This demonstrates a phase transition with symmetry reduction.
In contrast, the spectra at the $^{121}\nu_2(2)$ line showed a small anomaly at approximately 100 K.
No clear splits were observed.

\begin{figure}
    \centering
    \includegraphics{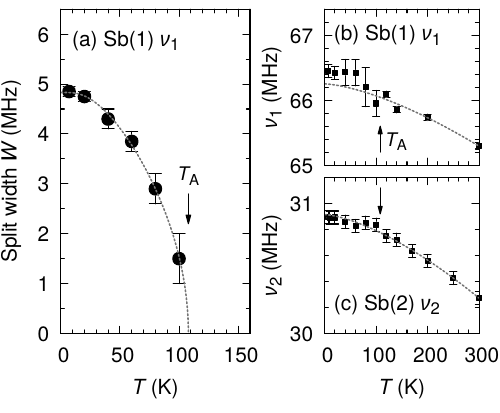}
    \caption{\label{fig:nuq-split}
        (a) Temperature dependence of the $^{121}$Sb NQR split width for Sb(1) site.
        The dashed line shows the fitting result with the BCS-like gap function obtained using the least-squares method.
        (b) The temperature dependence of the resonant frequency $\nu_{\textrm{1}}$ for Sb(1).
        (c) The temperature dependence of the resonant frequency $\nu_{2}$ for Sb(2).
        The dashed lines in (b) and (c) are the fitting result with an empirical
        temperature dependence $\nu_{\textrm{Q}}(T) = \nu_{\textrm{Q}}(0)(1-\alpha T^{3/2})$
        above $T_{\textrm{A}}$.
    }
\end{figure}

The site dependence of the NQR split can be explained by the
difference in the local structure between two sites.
Sb(1) site forms a quasi-one-dimensional chain along the $c$ axis possessing the CDW instability,
whereas Sb(2) forms a zigzag chain.
The nuclear relaxation shown later suggests the nonmagnetic transition.
Therefore,
the split peak in Fig.~\ref{fig:spectrum}(b) was analyzed using a spatially modulated quadrupole frequency
caused by the CDW transition
\begin{align}
    \nu_{\textrm{Q}}(x)=\nu_{0} + W\sin qx,\label{eq:nuqmodulation}
\end{align}
where $W$ and $q$ denote the amplitude and wavenumber of the modulation, respectively.
The nuclear position along the $\bm{q}$ vector is denoted by $x$.
If the wavelength $2\pi/q$ cannot be expressed as a rational multiple of the lattice constant,
$qx$ is uniformly distributed between zero and $2\pi$, which results in a continuous distribution of
$\nu_{\textrm{Q}}(x)$ between $\nu_{0}-W$ and $\nu_{0}+W$.
Then, the NQR intensity $I(f)$ at frequency $f$ is given by $I(f) df \propto d\theta$, where
$\theta = qx$ and $f=\nu_{0} + W\sin\theta$.
The NQR spectral shape is described as follows\cite{Phys.Rep.79.331}:
\begin{align}
    I(f) = \begin{cases}
               A\left[W^2 - (f-\nu_{0})^{2} \right]^{-1/2} & (\lvert f-\nu_{0} \rvert < W),\\
               0 & (\lvert f-\nu_{0} \rvert \ge W),
           \end{cases}\label{eq:spectralshape}
\end{align}
where $A$ denotes a constant.
This yields a double-peak spectrum at $f=\nu_{0}\pm W$ with a finite intensity between
two peaks.
The asymmetric parameter $\eta$ is fixed to zero for simplicity
in this model.
This can be determined by comparing frequencies between $^{121}\nu_1(1)$ and $^{121}\nu_2(1)$.

The fitting results obtained using Eq.~\eqref{eq:spectralshape} are represented by the dashed line in
Fig.~\ref{fig:spectrum}(b) for spectra at 100 K or below.
For a better fit to the experimental data, the above function was broadened using a Gaussian function
and is multiplied by a linearly varying weight function.
This simple model reproduces the experimental Sb(1) NQR spectrum at temperatures below 100 K.
The analyzed split width $W$ is shown in Fig.~\ref{fig:nuq-split}(a).
This temperature dependence appeared to be an order parameter with a second-order phase transition and
was analyzed using the following approximate BCS-type gap function\cite{PhysRevLett.84.5387}:
\begin{align}
    W(T)=W(0)\tanh\left[a(T_{\textrm{A}}/T-1)^{1/2} \right],
\end{align}
where $W(0)$, $a$, and $T_{\textrm{A}}$ denote fitting parameters.
Although the selection of this model is arbitrary, it approximately reproduces the experimental result,
as indicated by the dashed lines in Fig.~\ref{fig:nuq-split}(a).
The transition temperature was evaluated to be $T_{\textrm{A}} \simeq 110 $ K.
This value was close to 120 K, where an electrical resistivity anomaly was observed
in Ref.~\onlinecite{Chem.Mater.14.4867} and in the present study (Fig.~\ref{fig:resistivity}).
The relative split width regarding the quadrupole frequency is
$W(0)/\nu_{0} \simeq 0.073$ for $T \to 0$.
In contrast, the increase in the spectral width in $\nu_2(2)$ line is less than 0.15 MHz below $T_{\textrm{A}}$
estimated from the change in the half width at half maxima, yielding $W(0)/\nu_{0} \le 0.005$.
Because the NQR anomaly is more pronounced in Sb(1), which forms a quasi-one-dimensional chain,
this result supports the CDW transition originating from Fermi surface nesting.
The NQR result, combined with the electrical resistivity, was consistent with the
incommensurate CDW transition.

The anomaly at $T_{\textrm{A}}$ was also observed in the temperature dependence of $^{121}\nu_{1}(1)$
or $^{121}\nu_2(2)$ for Sb(1) and Sb(2), as shown in Fig.~\ref{fig:nuq-split}(b) and (c).
The frequencies gradually increased as the temperature decreased, which is ascribed to the lattice
shrinkage.
This can be explained by empirical temperature variations in
the NQR frequency\cite{ZPhysB.24.177}
$\nu_{\textrm{Q}}(T)=\nu_{\textrm{Q}}(0)(1-\alpha T^{3/2})$ above $T_{\textrm{A}}$,
and the fitted result is indicated by the dashed lines.
The deviation from this curve is pronounced in $^{121}\nu_{1}(1)$
below $T_{\textrm{A}}$, and a small kink is observed in $^{121}\nu_{2}(2)$ at $T_{\textrm{A}}$.
These results suggest a lattice constant change accompanying the CDW ordering.

\begin{figure}
    \centering
    \includegraphics{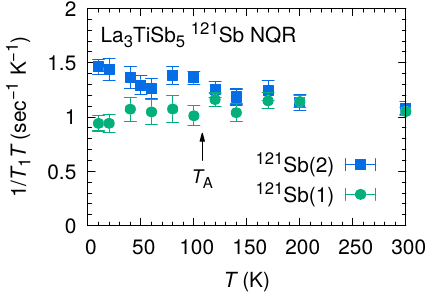}
    \caption{\label{fig:t1}
        (Color online) Nuclear spin-lattice relaxation rate divided by temperature ($1/T_1 T$)
        obtained by $^{121}$Sb NQR in La$_3$TiSb$_5$.
        The circle and square symbols represent the results at Sb(1) and Sb(2) sites, respectively.
        The vertical arrow shows the transition temperature determined by the Sb(1) line split.
    }
\end{figure}

Figure \ref{fig:t1} shows the nuclear spin-lattice relaxation rate divided by temperature  $1/T_1T$
for Sb(1) and Sb(2).
The relaxation rate probes the dynamics of the magnetic and electric origins.
The nuclear relaxation curve $M(t)$ was analyzed using
\begin{equation}
    \frac{M(\infty)-M(t)}{M(\infty)-M(0)}=R(t),
\end{equation}
where $R(t)$ denotes a normalized relaxation curve consisting of multiple exponential functions.
In the analysis, only the magnetic fluctuations were taken
into consideration for the nuclear relaxation process, and the quadrupole relaxation
mechanism was neglected.
$^{121}\nu_1(1)$ line corresponds to a $\pm 1/2 \leftrightarrow \pm 3/2$ transition and $R(t)$ is described as
\begin{equation}
    R(t)=\frac{3}{28} \exp\left(-\frac{3t}{T_1}\right) + \frac{25}{28} \exp\left(-\frac{10t}{T_1}\right).
\end{equation}
For $^{121}\nu_2(2)$, a finite $\eta$ altered the relaxation curve from that of the
$\pm 3/2 \leftrightarrow \pm 5/2$ transition, and thus, we adopt the following
function\cite{J.Phys.3.8103} obtained from numerical calculations for $\eta=0.38$:
\begin{equation}
    R(t)=0.3875\exp\left(-\frac{3t}{T_1}\right) + 0.6125\exp\left(-\frac{9.0875t}{T_1}\right).
\end{equation}
The relaxation curves exhibited a single component in two sites.
$1/T_1$ at the Sb(1) site was measured at the lower peaks below $T_{\textrm{A}}$.
Overall, $1/T_1T$ is nearly constant over the entire temperature range at these sites, indicating that
La$_3$TiSb$_5$ is a normal metal with small magnetic fluctuations.
Because magnetic fluctuations are absent in Sb(1), which exhibits an NQR line split,
the ordered phase is nonmagnetic.
This value is somewhat larger for Sb(2) than for Sb(1) below $T_{\textrm{A}}$,
which may be due to the change in the magnetic fluctuations
or hyperfine coupling constant caused by the ordering.

The absence of an anomaly in $1/T_1T$ may be unexpected, given the drastic NQR line split and resistivity hump.
The $1/T_1T$ behavior in the CDW state varies for different materials.
No detectable changes have been reported in the CDW state of 2H-NbSe$_2$\cite{JPSJ.65.2341} as in La$_3$TiSb$_5$,
whereas an enhancement of $1/T_1 T$ was observed in BaTi$_2$Sb$_2$O\cite{PhysRevB.87.060510}
and SrPt$_2$As$_2$\cite{PhysRevB.91.060510} toward $T_{\text{CDW}}$.
Site-dependent $1/T_1 T$ has been reported for Ta$_4$Pd$_3$Te$_{16}$\cite{PhysRevB.94.174511}
and CsV$_3$Sb$_5$\cite{Chin.Phys.Lett.38.077402,NPJ.Quantum.Mater.7.30}.
The small change in $^{121}$Sb $1/T_1T$ of La$_3$TiSb$_5$ suggests that $1/T_1T$
is mainly caused by the electrons not directly involved in the CDW transition,
and that the reduction of DOS accompanying the transition
does not strongly affect $1/T_1T$.
It is interesting to obtain $1/T_1T$ at the La site to clarify the site dependence
of the CDW anomaly.

Although the simple CDW model expressed in Eq.~\eqref{eq:nuqmodulation} reproduced the experimental results,
the NQR spectrum deviates slightly from the expected spectral shape in Eq.~\eqref{eq:spectralshape}:
the lower-frequency peaks were stronger, and an edge structure was observed at approximately 66 MHz at lower temperatures,
as shown in Fig.~\ref{fig:spectrum}(b).
Therefore, the actual CDW modulation may be more complicated than expected.
The complex spectrum may be related to the presence of Sb(1) and Ti chains with different symmetries.
% Existence of Sb(1) and Ti chains with different symmetries may relate to the complicated spectrum:
Sb(1) lacks local inversion symmetry, whereas Ti is located at the inversion center.
These factors can cause successive CDW transitions similar to those of
LaAgSb$_2$\cite{PhysRevB.68.035113} and SrPt$_2$As$_2$\cite{zaac.200700302,PhysRevB.85.184520,ChinPhysB.23.086103}.
Further studies, including x-ray or neutron diffraction experiments, may reveal the details of
the CDW structure in La$_3$TiSb$_5$.

Finally, we discuss the isostructural systems.
CDW instability may also exist in Ce$_3$AX$_5$ systems (A = Ti, Zr; X = Sb, Bi).
However, only antiferromagnetic transitions have been reported
thus far\cite{Chem.Mater.14.4867,J.Phys.Cond.Matt.29.145601,PhysicaB.536.142,JPSCP.30.011102,JPSCP.30.011180,J.Phys.2164.012040,JPSCP.38.011083}.
In Ce$_3$TiSb$_5$ and Ce$_3$TiBi$_5$, the magnetic structures are somewhat different.
Ce$_3$TiSb$_5$ exhibits a long-period modulation of the ordered moment along the easy axis
(in plane) with double transitions\cite{J.Phys.33.245801}.
In contrast, the cycloid structure has been reported for Ce$_3$TiBi$_5$, where the moment can be along the hard ($c$) axis\cite{PhysRevB.109.L140405}.
This difference may be partly due to the various mechanisms involved in the
magnetic structures of these systems.
The CDW instability inferred in this study can also be classified into these mechanisms.
Future experimental and theoretical studies will unravel the interplay between magnetism and CDW instability.

In summary, electrical resistivity and $^{121}$Sb NQR measurements were performed on
La$_3$TiSb$_5$, and the NQR spectrum of Sb(1) split below $T_{\text{A}} \simeq 110$ K.
This corresponds to a hump anomaly in the electrical resistivity.
The broad double-peak spectrum can be explained by the incommensurate CDW structure.
The nuclear spin-lattice relaxation rate shows no anomalies across the transition,
indicating the nonmagnetic ordering.
The CDW instability, owing to its quasi-one-dimensional character, can affect the magnetism of the systems
where the magnetoelectric effect is expected.

\begin{acknowledgments}
    The authors thank M. Yogi and T. Mutou for their insightful discussions.
    The authors also thank N. Masada and Y. Sano for their support in the experiments.
    This work was supported by JSPS KAKENHI Grant Numbers JP21K03447 and JP24K17012.
\end{acknowledgments}


%apsrev4-2.bst 2019-01-14 (MD) hand-edited version of apsrev4-1.bst
%Control: key (0)
%Control: author (8) initials jnrlst
%Control: editor formatted (1) identically to author
%Control: production of article title (0) allowed
%Control: page (0) single
%Control: year (1) truncated
%Control: production of eprint (0) enabled
\begin{thebibliography}{33}%
\makeatletter
\providecommand \@ifxundefined [1]{%
 \@ifx{#1\undefined}
}%
\providecommand \@ifnum [1]{%
 \ifnum #1\expandafter \@firstoftwo
 \else \expandafter \@secondoftwo
 \fi
}%
\providecommand \@ifx [1]{%
 \ifx #1\expandafter \@firstoftwo
 \else \expandafter \@secondoftwo
 \fi
}%
\providecommand \natexlab [1]{#1}%
\providecommand \enquote  [1]{``#1''}%
\providecommand \bibnamefont  [1]{#1}%
\providecommand \bibfnamefont [1]{#1}%
\providecommand \citenamefont [1]{#1}%
\providecommand \href@noop [0]{\@secondoftwo}%
\providecommand \href [0]{\begingroup \@sanitize@url \@href}%
\providecommand \@href[1]{\@@startlink{#1}\@@href}%
\providecommand \@@href[1]{\endgroup#1\@@endlink}%
\providecommand \@sanitize@url [0]{\catcode `\\12\catcode `\$12\catcode
  `\&12\catcode `\#12\catcode `\^12\catcode `\_12\catcode `\%12\relax}%
\providecommand \@@startlink[1]{}%
\providecommand \@@endlink[0]{}%
\providecommand \url  [0]{\begingroup\@sanitize@url \@url }%
\providecommand \@url [1]{\endgroup\@href {#1}{\urlprefix }}%
\providecommand \urlprefix  [0]{URL }%
\providecommand \Eprint [0]{\href }%
\providecommand \doibase [0]{https://doi.org/}%
\providecommand \selectlanguage [0]{\@gobble}%
\providecommand \bibinfo  [0]{\@secondoftwo}%
\providecommand \bibfield  [0]{\@secondoftwo}%
\providecommand \translation [1]{[#1]}%
\providecommand \BibitemOpen [0]{}%
\providecommand \bibitemStop [0]{}%
\providecommand \bibitemNoStop [0]{.\EOS\space}%
\providecommand \EOS [0]{\spacefactor3000\relax}%
\providecommand \BibitemShut  [1]{\csname bibitem#1\endcsname}%
\let\auto@bib@innerbib\@empty
%</preamble>
\bibitem [{\citenamefont {Sumita}\ and\ \citenamefont
  {Yanase}(2016)}]{PhysRevB.93.224507}%
  \BibitemOpen
  \bibfield  {author} {\bibinfo {author} {\bibfnamefont {S.}~\bibnamefont
  {Sumita}}\ and\ \bibinfo {author} {\bibfnamefont {Y.}~\bibnamefont
  {Yanase}},\ }\bibfield  {title} {\bibinfo {title} {Superconductivity in
  magnetic multipole states},\ }\href
  {https://doi.org/10.1103/PhysRevB.93.224507} {\bibfield  {journal} {\bibinfo
  {journal} {Phys. Rev. B}\ }\textbf {\bibinfo {volume} {93}},\ \bibinfo
  {pages} {224507} (\bibinfo {year} {2016})}\BibitemShut {NoStop}%
\bibitem [{\citenamefont {Ishizuka}\ and\ \citenamefont
  {Yanase}(2018)}]{PhysRevB.98.224510}%
  \BibitemOpen
  \bibfield  {author} {\bibinfo {author} {\bibfnamefont {J.}~\bibnamefont
  {Ishizuka}}\ and\ \bibinfo {author} {\bibfnamefont {Y.}~\bibnamefont
  {Yanase}},\ }\bibfield  {title} {\bibinfo {title} {Odd-parity multipole
  fluctuation and unconventional superconductivity in locally
  noncentrosymmetric crystal},\ }\href
  {https://doi.org/10.1103/PhysRevB.98.224510} {\bibfield  {journal} {\bibinfo
  {journal} {Phys. Rev. B}\ }\textbf {\bibinfo {volume} {98}},\ \bibinfo
  {pages} {224510} (\bibinfo {year} {2018})}\BibitemShut {NoStop}%
\bibitem [{\citenamefont {Hayami}\ \emph {et~al.}(2014)\citenamefont {Hayami},
  \citenamefont {Kusunose},\ and\ \citenamefont {Motome}}]{PhysRevB.90.024432}%
  \BibitemOpen
  \bibfield  {author} {\bibinfo {author} {\bibfnamefont {S.}~\bibnamefont
  {Hayami}}, \bibinfo {author} {\bibfnamefont {H.}~\bibnamefont {Kusunose}},\
  and\ \bibinfo {author} {\bibfnamefont {Y.}~\bibnamefont {Motome}},\
  }\bibfield  {title} {\bibinfo {title} {Toroidal order in metals without local
  inversion symmetry},\ }\href {https://doi.org/10.1103/PhysRevB.90.024432}
  {\bibfield  {journal} {\bibinfo  {journal} {Phys. Rev. B}\ }\textbf {\bibinfo
  {volume} {90}},\ \bibinfo {pages} {024432} (\bibinfo {year}
  {2014})}\BibitemShut {NoStop}%
\bibitem [{\citenamefont {Yanase}(2014)}]{JPSJ.83.014703}%
  \BibitemOpen
  \bibfield  {author} {\bibinfo {author} {\bibfnamefont {Y.}~\bibnamefont
  {Yanase}},\ }\bibfield  {title} {\bibinfo {title} {Magneto-electric effect in
  three-dimensional coupled zigzag chains},\ }\href
  {https://doi.org/10.7566/JPSJ.83.014703} {\bibfield  {journal} {\bibinfo
  {journal} {J. Phys. Soc. Jpn.}\ }\textbf {\bibinfo {volume} {83}},\ \bibinfo
  {pages} {014703} (\bibinfo {year} {2014})}\BibitemShut {NoStop}%
\bibitem [{\citenamefont {Saito}\ \emph {et~al.}(2018)\citenamefont {Saito},
  \citenamefont {Uenishi}, \citenamefont {Miura}, \citenamefont {Tabata},
  \citenamefont {Hidaka}, \citenamefont {Yanagisawa},\ and\ \citenamefont
  {Amitsuka}}]{JPSJ.87.033702}%
  \BibitemOpen
  \bibfield  {author} {\bibinfo {author} {\bibfnamefont {H.}~\bibnamefont
  {Saito}}, \bibinfo {author} {\bibfnamefont {K.}~\bibnamefont {Uenishi}},
  \bibinfo {author} {\bibfnamefont {N.}~\bibnamefont {Miura}}, \bibinfo
  {author} {\bibfnamefont {C.}~\bibnamefont {Tabata}}, \bibinfo {author}
  {\bibfnamefont {H.}~\bibnamefont {Hidaka}}, \bibinfo {author} {\bibfnamefont
  {T.}~\bibnamefont {Yanagisawa}},\ and\ \bibinfo {author} {\bibfnamefont
  {H.}~\bibnamefont {Amitsuka}},\ }\bibfield  {title} {\bibinfo {title}
  {Evidence of a new current-induced magnetoelectric effect in a toroidal
  magnetic ordered state of {UNi$_4$B}},\ }\href
  {https://doi.org/10.7566/JPSJ.87.033702} {\bibfield  {journal} {\bibinfo
  {journal} {J. Phys. Soc. Jpn.}\ }\textbf {\bibinfo {volume} {87}},\ \bibinfo
  {pages} {033702} (\bibinfo {year} {2018})}\BibitemShut {NoStop}%
\bibitem [{\citenamefont {Shinozaki}\ \emph
  {et~al.}(2020{\natexlab{a}})\citenamefont {Shinozaki}, \citenamefont
  {Motoyama}, \citenamefont {Tsubouchi}, \citenamefont {Sezaki}, \citenamefont
  {Gouchi}, \citenamefont {Nishigori}, \citenamefont {Mutou}, \citenamefont
  {Yamaguchi}, \citenamefont {Fujiwara}, \citenamefont {Miyoshi},\ and\
  \citenamefont {Uwatoko}}]{JPSJ.89.033703}%
  \BibitemOpen
  \bibfield  {author} {\bibinfo {author} {\bibfnamefont {M.}~\bibnamefont
  {Shinozaki}}, \bibinfo {author} {\bibfnamefont {G.}~\bibnamefont {Motoyama}},
  \bibinfo {author} {\bibfnamefont {M.}~\bibnamefont {Tsubouchi}}, \bibinfo
  {author} {\bibfnamefont {M.}~\bibnamefont {Sezaki}}, \bibinfo {author}
  {\bibfnamefont {J.}~\bibnamefont {Gouchi}}, \bibinfo {author} {\bibfnamefont
  {S.}~\bibnamefont {Nishigori}}, \bibinfo {author} {\bibfnamefont
  {T.}~\bibnamefont {Mutou}}, \bibinfo {author} {\bibfnamefont
  {A.}~\bibnamefont {Yamaguchi}}, \bibinfo {author} {\bibfnamefont
  {K.}~\bibnamefont {Fujiwara}}, \bibinfo {author} {\bibfnamefont
  {K.}~\bibnamefont {Miyoshi}},\ and\ \bibinfo {author} {\bibfnamefont
  {Y.}~\bibnamefont {Uwatoko}},\ }\bibfield  {title} {\bibinfo {title}
  {Magnetoelectric effect in the antiferromagnetic ordered state of
  {Ce}$_3${TiBi}$_5$ with {Ce} zig-zag chains},\ }\href
  {https://doi.org/10.7566/JPSJ.89.033703} {\bibfield  {journal} {\bibinfo
  {journal} {J. Phys. Soc. Jpn.}\ }\textbf {\bibinfo {volume} {89}},\ \bibinfo
  {pages} {033703} (\bibinfo {year} {2020}{\natexlab{a}})}\BibitemShut
  {NoStop}%
\bibitem [{\citenamefont {Shinozaki}\ \emph
  {et~al.}(2020{\natexlab{b}})\citenamefont {Shinozaki}, \citenamefont
  {Motoyama}, \citenamefont {Mutou}, \citenamefont {Nishigori}, \citenamefont
  {Yamaguchi}, \citenamefont {Fujiwara}, \citenamefont {Miyoshi},\ and\
  \citenamefont {Sumiyama}}]{JPSCP.30.011189}%
  \BibitemOpen
  \bibfield  {author} {\bibinfo {author} {\bibfnamefont {M.}~\bibnamefont
  {Shinozaki}}, \bibinfo {author} {\bibfnamefont {G.}~\bibnamefont {Motoyama}},
  \bibinfo {author} {\bibfnamefont {T.}~\bibnamefont {Mutou}}, \bibinfo
  {author} {\bibfnamefont {S.}~\bibnamefont {Nishigori}}, \bibinfo {author}
  {\bibfnamefont {A.}~\bibnamefont {Yamaguchi}}, \bibinfo {author}
  {\bibfnamefont {K.}~\bibnamefont {Fujiwara}}, \bibinfo {author}
  {\bibfnamefont {K.}~\bibnamefont {Miyoshi}},\ and\ \bibinfo {author}
  {\bibfnamefont {A.}~\bibnamefont {Sumiyama}},\ }\bibfield  {title} {\bibinfo
  {title} {Study for current-induced magnetization in ferrotoroidal ordered
  state of {Ce$_3$TiBi$_5$}},\ }\bibfield  {booktitle} {\emph {\bibinfo
  {booktitle} {Proceedings of the International Conference on Strongly
  Correlated Electron Systems (SCES2019)}},\ }\href
  {https://doi.org/10.7566/JPSCP.30.011189} {\bibfield  {journal} {\bibinfo
  {journal} {JPS Conf. Proc.}\ }\textbf {\bibinfo {volume} {30}},\ \bibinfo
  {pages} {011189} (\bibinfo {year} {2020}{\natexlab{b}})}\BibitemShut
  {NoStop}%
\bibitem [{\citenamefont {Bolloré}\ \emph {et~al.}(1995)\citenamefont
  {Bolloré}, \citenamefont {Ferguson}, \citenamefont {Hushagen},\ and\
  \citenamefont {Mar}}]{Chem.Mater.7.2229}%
  \BibitemOpen
  \bibfield  {author} {\bibinfo {author} {\bibfnamefont {G.}~\bibnamefont
  {Bolloré}}, \bibinfo {author} {\bibfnamefont {M.~J.}\ \bibnamefont
  {Ferguson}}, \bibinfo {author} {\bibfnamefont {R.~W.}\ \bibnamefont
  {Hushagen}},\ and\ \bibinfo {author} {\bibfnamefont {A.}~\bibnamefont
  {Mar}},\ }\bibfield  {title} {\bibinfo {title} {New ternary rare-earth
  transition-metal antimonides {RE$_3$MSb$_5$} ({RE = La, Ce, Pr, Nd, Sm; M =
  Ti, Zr, Hf, Nb})},\ }\href {https://doi.org/10.1021/cm00060a005} {\bibfield
  {journal} {\bibinfo  {journal} {Chem. Mater.}\ }\textbf {\bibinfo {volume}
  {7}},\ \bibinfo {pages} {2229} (\bibinfo {year} {1995})}\BibitemShut
  {NoStop}%
\bibitem [{\citenamefont {Motoyama}\ \emph {et~al.}(2018)\citenamefont
  {Motoyama}, \citenamefont {Sezaki}, \citenamefont {Gouchi}, \citenamefont
  {Miyoshi}, \citenamefont {Nishigori}, \citenamefont {Mutou}, \citenamefont
  {Fujiwara},\ and\ \citenamefont {Uwatoko}}]{PhysicaB.536.142}%
  \BibitemOpen
  \bibfield  {author} {\bibinfo {author} {\bibfnamefont {G.}~\bibnamefont
  {Motoyama}}, \bibinfo {author} {\bibfnamefont {M.}~\bibnamefont {Sezaki}},
  \bibinfo {author} {\bibfnamefont {J.}~\bibnamefont {Gouchi}}, \bibinfo
  {author} {\bibfnamefont {K.}~\bibnamefont {Miyoshi}}, \bibinfo {author}
  {\bibfnamefont {S.}~\bibnamefont {Nishigori}}, \bibinfo {author}
  {\bibfnamefont {T.}~\bibnamefont {Mutou}}, \bibinfo {author} {\bibfnamefont
  {K.}~\bibnamefont {Fujiwara}},\ and\ \bibinfo {author} {\bibfnamefont
  {Y.}~\bibnamefont {Uwatoko}},\ }\bibfield  {title} {\bibinfo {title}
  {Magnetic properties of new antiferromagnetic heavy-fermion compounds,
  {Ce$_3$TiBi$_5$} and {CeTi$_3$Bi$_4$}},\ }\href
  {https://doi.org/10.1016/j.physb.2017.10.005} {\bibfield  {journal} {\bibinfo
   {journal} {Physica B: Condensed Matter}\ }\textbf {\bibinfo {volume}
  {536}},\ \bibinfo {pages} {142} (\bibinfo {year} {2018})}\BibitemShut
  {NoStop}%
\bibitem [{\citenamefont {Gauthier}\ \emph {et~al.}(2024)\citenamefont
  {Gauthier}, \citenamefont {Sibille}, \citenamefont {Pomjakushin},
  \citenamefont {Fjellv\aa{}g}, \citenamefont {Fraser}, \citenamefont
  {Desmarais}, \citenamefont {Bianchi},\ and\ \citenamefont
  {Quilliam}}]{PhysRevB.109.L140405}%
  \BibitemOpen
  \bibfield  {author} {\bibinfo {author} {\bibfnamefont {N.}~\bibnamefont
  {Gauthier}}, \bibinfo {author} {\bibfnamefont {R.}~\bibnamefont {Sibille}},
  \bibinfo {author} {\bibfnamefont {V.}~\bibnamefont {Pomjakushin}}, \bibinfo
  {author} {\bibfnamefont {O.~S.}\ \bibnamefont {Fjellv\aa{}g}}, \bibinfo
  {author} {\bibfnamefont {J.}~\bibnamefont {Fraser}}, \bibinfo {author}
  {\bibfnamefont {M.}~\bibnamefont {Desmarais}}, \bibinfo {author}
  {\bibfnamefont {A.~D.}\ \bibnamefont {Bianchi}},\ and\ \bibinfo {author}
  {\bibfnamefont {J.~A.}\ \bibnamefont {Quilliam}},\ }\bibfield  {title}
  {\bibinfo {title} {Magnetic structure of
  {${\mathrm{Ce}}_{3}{\mathrm{TiBi}}_{5}$} and its relation to current-induced
  magnetization},\ }\href {https://doi.org/10.1103/PhysRevB.109.L140405}
  {\bibfield  {journal} {\bibinfo  {journal} {Phys. Rev. B}\ }\textbf {\bibinfo
  {volume} {109}},\ \bibinfo {pages} {L140405} (\bibinfo {year}
  {2024})}\BibitemShut {NoStop}%
\bibitem [{\citenamefont {Ritter}\ \emph {et~al.}(2021)\citenamefont {Ritter},
  \citenamefont {Pathak}, \citenamefont {Filippone}, \citenamefont {Provino},
  \citenamefont {Dhar},\ and\ \citenamefont {Manfrinetti}}]{J.Phys.33.245801}%
  \BibitemOpen
  \bibfield  {author} {\bibinfo {author} {\bibfnamefont {C.}~\bibnamefont
  {Ritter}}, \bibinfo {author} {\bibfnamefont {A.~K.}\ \bibnamefont {Pathak}},
  \bibinfo {author} {\bibfnamefont {R.}~\bibnamefont {Filippone}}, \bibinfo
  {author} {\bibfnamefont {A.}~\bibnamefont {Provino}}, \bibinfo {author}
  {\bibfnamefont {S.~K.}\ \bibnamefont {Dhar}},\ and\ \bibinfo {author}
  {\bibfnamefont {P.}~\bibnamefont {Manfrinetti}},\ }\bibfield  {title}
  {\bibinfo {title} {Magnetic ground states of {Ce$_3$TiSb$_5$},
  {Pr$_3$TiSb$_5$} and {Nd$_3$TiSb$_5$} determined by neutron powder
  diffraction and magnetic measurements},\ }\href
  {https://doi.org/10.1088/1361-648X/abe9db} {\bibfield  {journal} {\bibinfo
  {journal} {J. Phys.: Cond. Matter}\ }\textbf {\bibinfo {volume} {33}},\
  \bibinfo {pages} {245801} (\bibinfo {year} {2021})}\BibitemShut {NoStop}%
\bibitem [{\citenamefont {Moore}\ \emph {et~al.}(2002)\citenamefont {Moore},
  \citenamefont {Deakin}, \citenamefont {Ferguson},\ and\ \citenamefont
  {Mar}}]{Chem.Mater.14.4867}%
  \BibitemOpen
  \bibfield  {author} {\bibinfo {author} {\bibfnamefont {S.~H.~D.}\
  \bibnamefont {Moore}}, \bibinfo {author} {\bibfnamefont {L.}~\bibnamefont
  {Deakin}}, \bibinfo {author} {\bibfnamefont {M.~J.}\ \bibnamefont
  {Ferguson}},\ and\ \bibinfo {author} {\bibfnamefont {A.}~\bibnamefont
  {Mar}},\ }\bibfield  {title} {\bibinfo {title} {Physical properties and
  bonding in {RE$_3$TiSb$_5$} ({RE = La, Ce, Pr, Nd, Sm})},\ }\href
  {https://doi.org/10.1021/cm020731t} {\bibfield  {journal} {\bibinfo
  {journal} {Chem. Mater.}\ }\textbf {\bibinfo {volume} {14}},\ \bibinfo
  {pages} {4867} (\bibinfo {year} {2002})}\BibitemShut {NoStop}%
\bibitem [{\citenamefont {Momma}\ and\ \citenamefont
  {Izumi}(2011)}]{JApplCryst.44.1272}%
  \BibitemOpen
  \bibfield  {author} {\bibinfo {author} {\bibfnamefont {K.}~\bibnamefont
  {Momma}}\ and\ \bibinfo {author} {\bibfnamefont {F.}~\bibnamefont {Izumi}},\
  }\bibfield  {title} {\bibinfo {title} {{{\textit{VESTA3}} for
  three-dimensional visualization of crystal, volumetric and morphology
  data}},\ }\href {https://doi.org/10.1107/S0021889811038970} {\bibfield
  {journal} {\bibinfo  {journal} {J. Appl. Cryst.}\ }\textbf {\bibinfo {volume}
  {44}},\ \bibinfo {pages} {1272} (\bibinfo {year} {2011})}\BibitemShut
  {NoStop}%
\bibitem [{\citenamefont {Matin}\ \emph {et~al.}(2017)\citenamefont {Matin},
  \citenamefont {Kulkarni}, \citenamefont {Thamizhavel}, \citenamefont {Dhar},
  \citenamefont {Provino},\ and\ \citenamefont
  {Manfrinetti}}]{J.Phys.Cond.Matt.29.145601}%
  \BibitemOpen
  \bibfield  {author} {\bibinfo {author} {\bibfnamefont {M.}~\bibnamefont
  {Matin}}, \bibinfo {author} {\bibfnamefont {R.}~\bibnamefont {Kulkarni}},
  \bibinfo {author} {\bibfnamefont {A.}~\bibnamefont {Thamizhavel}}, \bibinfo
  {author} {\bibfnamefont {S.~K.}\ \bibnamefont {Dhar}}, \bibinfo {author}
  {\bibfnamefont {A.}~\bibnamefont {Provino}},\ and\ \bibinfo {author}
  {\bibfnamefont {P.}~\bibnamefont {Manfrinetti}},\ }\bibfield  {title}
  {\bibinfo {title} {Probing the magnetic ground state of single crystalline
  {Ce$_3$TiSb$_5$}},\ }\href {https://doi.org/10.1088/1361-648X/aa57c0}
  {\bibfield  {journal} {\bibinfo  {journal} {J. Phys: Cond. Matter}\ }\textbf
  {\bibinfo {volume} {29}},\ \bibinfo {pages} {145601} (\bibinfo {year}
  {2017})}\BibitemShut {NoStop}%
\bibitem [{\citenamefont {Stone}(2016)}]{AtData.111.1}%
  \BibitemOpen
  \bibfield  {author} {\bibinfo {author} {\bibfnamefont {N.}~\bibnamefont
  {Stone}},\ }\bibfield  {title} {\bibinfo {title} {Table of nuclear electric
  quadrupole moments},\ }\href {https://doi.org/10.1016/j.adt.2015.12.002}
  {\bibfield  {journal} {\bibinfo  {journal} {At. Data Nucl. Data Tables}\
  }\textbf {\bibinfo {volume} {111-112}},\ \bibinfo {pages} {1 } (\bibinfo
  {year} {2016})}\BibitemShut {NoStop}%
\bibitem [{\citenamefont {Blinc}(1981)}]{Phys.Rep.79.331}%
  \BibitemOpen
  \bibfield  {author} {\bibinfo {author} {\bibfnamefont {R.}~\bibnamefont
  {Blinc}},\ }\bibfield  {title} {\bibinfo {title} {Magnetic resonance and
  relaxation in structurally incommensurate systems},\ }\href
  {https://doi.org/10.1016/0370-1573(81)90108-3} {\bibfield  {journal}
  {\bibinfo  {journal} {Phys. Rep.}\ }\textbf {\bibinfo {volume} {79}},\
  \bibinfo {pages} {331} (\bibinfo {year} {1981})}\BibitemShut {NoStop}%
\bibitem [{\citenamefont {Ishida}\ \emph {et~al.}(2000)\citenamefont {Ishida},
  \citenamefont {Mukuda}, \citenamefont {Kitaoka}, \citenamefont {Mao},
  \citenamefont {Mori},\ and\ \citenamefont {Maeno}}]{PhysRevLett.84.5387}%
  \BibitemOpen
  \bibfield  {author} {\bibinfo {author} {\bibfnamefont {K.}~\bibnamefont
  {Ishida}}, \bibinfo {author} {\bibfnamefont {H.}~\bibnamefont {Mukuda}},
  \bibinfo {author} {\bibfnamefont {Y.}~\bibnamefont {Kitaoka}}, \bibinfo
  {author} {\bibfnamefont {Z.~Q.}\ \bibnamefont {Mao}}, \bibinfo {author}
  {\bibfnamefont {Y.}~\bibnamefont {Mori}},\ and\ \bibinfo {author}
  {\bibfnamefont {Y.}~\bibnamefont {Maeno}},\ }\bibfield  {title} {\bibinfo
  {title} {Anisotropic superconducting gap in the spin-triplet superconductor
  {${\mathrm{Sr}}_{2}{\mathrm{RuO}}_{4}$}: Evidence from a {Ru-NQR} study},\
  }\href {https://doi.org/10.1103/PhysRevLett.84.5387} {\bibfield  {journal}
  {\bibinfo  {journal} {Phys. Rev. Lett.}\ }\textbf {\bibinfo {volume} {84}},\
  \bibinfo {pages} {5387} (\bibinfo {year} {2000})}\BibitemShut {NoStop}%
\bibitem [{\citenamefont {Christiansen}\ \emph {et~al.}(1976)\citenamefont
  {Christiansen}, \citenamefont {Heubes}, \citenamefont {Keitel}, \citenamefont
  {Klinger}, \citenamefont {Loeffler}, \citenamefont {Sandner},\ and\
  \citenamefont {Witthuhn}}]{ZPhysB.24.177}%
  \BibitemOpen
  \bibfield  {author} {\bibinfo {author} {\bibfnamefont {J.}~\bibnamefont
  {Christiansen}}, \bibinfo {author} {\bibfnamefont {P.}~\bibnamefont
  {Heubes}}, \bibinfo {author} {\bibfnamefont {R.}~\bibnamefont {Keitel}},
  \bibinfo {author} {\bibfnamefont {W.}~\bibnamefont {Klinger}}, \bibinfo
  {author} {\bibfnamefont {W.}~\bibnamefont {Loeffler}}, \bibinfo {author}
  {\bibfnamefont {W.}~\bibnamefont {Sandner}},\ and\ \bibinfo {author}
  {\bibfnamefont {W.}~\bibnamefont {Witthuhn}},\ }\bibfield  {title} {\bibinfo
  {title} {Temperature dependence of the electric field gradient in noncubic
  metals},\ }\href {https://doi.org/10.1007/BF01312998} {\bibfield  {journal}
  {\bibinfo  {journal} {Z. Phys. B}\ }\textbf {\bibinfo {volume} {24}},\
  \bibinfo {pages} {177} (\bibinfo {year} {1976})}\BibitemShut {NoStop}%
\bibitem [{\citenamefont {Chepin}\ and\ \citenamefont
  {Ross}(1991)}]{J.Phys.3.8103}%
  \BibitemOpen
  \bibfield  {author} {\bibinfo {author} {\bibfnamefont {J.}~\bibnamefont
  {Chepin}}\ and\ \bibinfo {author} {\bibfnamefont {J.~H.}\ \bibnamefont
  {Ross}, \bibfnamefont {Jr.}},\ }\bibfield  {title} {\bibinfo {title}
  {Magnetic spin-lattice relaxation in nuclear quadrupole resonance: the eta
  not=0 case},\ }\href {https://doi.org/10.1088/0953-8984/3/41/009} {\bibfield
  {journal} {\bibinfo  {journal} {J. Phys.: Condens. Matter}\ }\textbf
  {\bibinfo {volume} {3}},\ \bibinfo {pages} {8103} (\bibinfo {year}
  {1991})}\BibitemShut {NoStop}%
\bibitem [{\citenamefont {Ishida}\ \emph {et~al.}(1996)\citenamefont {Ishida},
  \citenamefont {Niino}, \citenamefont {Zheng}, \citenamefont {Kitaoka},
  \citenamefont {Asayama},\ and\ \citenamefont {Ohtani}}]{JPSJ.65.2341}%
  \BibitemOpen
  \bibfield  {author} {\bibinfo {author} {\bibfnamefont {K.}~\bibnamefont
  {Ishida}}, \bibinfo {author} {\bibfnamefont {Y.}~\bibnamefont {Niino}},
  \bibinfo {author} {\bibfnamefont {G.-Q.}\ \bibnamefont {Zheng}}, \bibinfo
  {author} {\bibfnamefont {Y.}~\bibnamefont {Kitaoka}}, \bibinfo {author}
  {\bibfnamefont {K.}~\bibnamefont {Asayama}},\ and\ \bibinfo {author}
  {\bibfnamefont {T.}~\bibnamefont {Ohtani}},\ }\bibfield  {title} {\bibinfo
  {title} {$^{93}${Nb NQR} study in layered superconducting {2H-NbSe$_2$}},\
  }\href {https://doi.org/10.1143/JPSJ.65.2341} {\bibfield  {journal} {\bibinfo
   {journal} {J. Phys. Soc. Jpn.}\ }\textbf {\bibinfo {volume} {65}},\ \bibinfo
  {pages} {2341} (\bibinfo {year} {1996})}\BibitemShut {NoStop}%
\bibitem [{\citenamefont {Kitagawa}\ \emph {et~al.}(2013)\citenamefont
  {Kitagawa}, \citenamefont {Ishida}, \citenamefont {Nakano}, \citenamefont
  {Yajima},\ and\ \citenamefont {Kageyama}}]{PhysRevB.87.060510}%
  \BibitemOpen
  \bibfield  {author} {\bibinfo {author} {\bibfnamefont {S.}~\bibnamefont
  {Kitagawa}}, \bibinfo {author} {\bibfnamefont {K.}~\bibnamefont {Ishida}},
  \bibinfo {author} {\bibfnamefont {K.}~\bibnamefont {Nakano}}, \bibinfo
  {author} {\bibfnamefont {T.}~\bibnamefont {Yajima}},\ and\ \bibinfo {author}
  {\bibfnamefont {H.}~\bibnamefont {Kageyama}},\ }\bibfield  {title} {\bibinfo
  {title} {$s$-wave superconductivity in superconducting
  {BaTi${}_{2}$Sb${}_{2}$O} revealed by {${}^{121/123}$Sb-NMR}/nuclear
  quadrupole resonance measurements},\ }\href
  {https://doi.org/10.1103/PhysRevB.87.060510} {\bibfield  {journal} {\bibinfo
  {journal} {Phys. Rev. B}\ }\textbf {\bibinfo {volume} {87}},\ \bibinfo
  {pages} {060510} (\bibinfo {year} {2013})}\BibitemShut {NoStop}%
\bibitem [{\citenamefont {Kawasaki}\ \emph {et~al.}(2015)\citenamefont
  {Kawasaki}, \citenamefont {Tani}, \citenamefont {Mabuchi}, \citenamefont
  {Kudo}, \citenamefont {Nishikubo}, \citenamefont {Mitsuoka}, \citenamefont
  {Nohara},\ and\ \citenamefont {Zheng}}]{PhysRevB.91.060510}%
  \BibitemOpen
  \bibfield  {author} {\bibinfo {author} {\bibfnamefont {S.}~\bibnamefont
  {Kawasaki}}, \bibinfo {author} {\bibfnamefont {Y.}~\bibnamefont {Tani}},
  \bibinfo {author} {\bibfnamefont {T.}~\bibnamefont {Mabuchi}}, \bibinfo
  {author} {\bibfnamefont {K.}~\bibnamefont {Kudo}}, \bibinfo {author}
  {\bibfnamefont {Y.}~\bibnamefont {Nishikubo}}, \bibinfo {author}
  {\bibfnamefont {D.}~\bibnamefont {Mitsuoka}}, \bibinfo {author}
  {\bibfnamefont {M.}~\bibnamefont {Nohara}},\ and\ \bibinfo {author}
  {\bibfnamefont {G.-q.}\ \bibnamefont {Zheng}},\ }\bibfield  {title} {\bibinfo
  {title} {Coexistence of multiple charge-density waves and superconductivity
  in {${\text{SrPt}}_{2}{\text{As}}_{2}$} revealed by
  {$^{75}\mathrm{As}\ensuremath{-}\mathrm{NMR}/\mathrm{NQR}$} and
  {$^{195}\mathrm{Pt}\ensuremath{-}\mathrm{NMR}$}},\ }\href
  {https://doi.org/10.1103/PhysRevB.91.060510} {\bibfield  {journal} {\bibinfo
  {journal} {Phys. Rev. B}\ }\textbf {\bibinfo {volume} {91}},\ \bibinfo
  {pages} {060510} (\bibinfo {year} {2015})}\BibitemShut {NoStop}%
\bibitem [{\citenamefont {Li}\ \emph {et~al.}(2016)\citenamefont {Li},
  \citenamefont {Jiao}, \citenamefont {Cao},\ and\ \citenamefont
  {Zheng}}]{PhysRevB.94.174511}%
  \BibitemOpen
  \bibfield  {author} {\bibinfo {author} {\bibfnamefont {Z.}~\bibnamefont
  {Li}}, \bibinfo {author} {\bibfnamefont {W.~H.}\ \bibnamefont {Jiao}},
  \bibinfo {author} {\bibfnamefont {G.~H.}\ \bibnamefont {Cao}},\ and\ \bibinfo
  {author} {\bibfnamefont {G.-q.}\ \bibnamefont {Zheng}},\ }\bibfield  {title}
  {\bibinfo {title} {Charge fluctuations and nodeless superconductivity in
  quasi-one-dimensional {Ta}$_{4}${Pd}$_{3}${Te}$_{16}$ revealed by
  {$^{125}$Te-NMR} and {$^{181}$Ta-NQR}},\ }\href
  {https://doi.org/10.1103/PhysRevB.94.174511} {\bibfield  {journal} {\bibinfo
  {journal} {Phys. Rev. B}\ }\textbf {\bibinfo {volume} {94}},\ \bibinfo
  {pages} {174511} (\bibinfo {year} {2016})}\BibitemShut {NoStop}%
\bibitem [{\citenamefont {Mu}\ \emph {et~al.}(2021)\citenamefont {Mu},
  \citenamefont {Yin}, \citenamefont {Tu}, \citenamefont {Gong}, \citenamefont
  {Lei}, \citenamefont {Li},\ and\ \citenamefont
  {Luo}}]{Chin.Phys.Lett.38.077402}%
  \BibitemOpen
  \bibfield  {author} {\bibinfo {author} {\bibfnamefont {C.}~\bibnamefont
  {Mu}}, \bibinfo {author} {\bibfnamefont {Q.}~\bibnamefont {Yin}}, \bibinfo
  {author} {\bibfnamefont {Z.}~\bibnamefont {Tu}}, \bibinfo {author}
  {\bibfnamefont {C.}~\bibnamefont {Gong}}, \bibinfo {author} {\bibfnamefont
  {H.}~\bibnamefont {Lei}}, \bibinfo {author} {\bibfnamefont {Z.}~\bibnamefont
  {Li}},\ and\ \bibinfo {author} {\bibfnamefont {J.}~\bibnamefont {Luo}},\
  }\bibfield  {title} {\bibinfo {title} {S-wave superconductivity in kagome
  metal {CsV$_3$Sb$_5$} revealed by {$^{121/123}$Sb} {NQR} and {$^{51}$V NMR}
  measurements},\ }\href {https://doi.org/10.1088/0256-307X/38/7/077402}
  {\bibfield  {journal} {\bibinfo  {journal} {Chinese Phys. Lett.}\ }\textbf
  {\bibinfo {volume} {38}},\ \bibinfo {pages} {077402} (\bibinfo {year}
  {2021})}\BibitemShut {NoStop}%
\bibitem [{\citenamefont {Luo}\ \emph {et~al.}(2022)\citenamefont {Luo},
  \citenamefont {Zhao}, \citenamefont {Zhou}, \citenamefont {Yang},
  \citenamefont {Fang}, \citenamefont {Yang}, \citenamefont {Gao},
  \citenamefont {Zhou},\ and\ \citenamefont {Zheng}}]{NPJ.Quantum.Mater.7.30}%
  \BibitemOpen
  \bibfield  {author} {\bibinfo {author} {\bibfnamefont {J.}~\bibnamefont
  {Luo}}, \bibinfo {author} {\bibfnamefont {Z.}~\bibnamefont {Zhao}}, \bibinfo
  {author} {\bibfnamefont {Y.~Z.}\ \bibnamefont {Zhou}}, \bibinfo {author}
  {\bibfnamefont {J.}~\bibnamefont {Yang}}, \bibinfo {author} {\bibfnamefont
  {A.~F.}\ \bibnamefont {Fang}}, \bibinfo {author} {\bibfnamefont {H.~T.}\
  \bibnamefont {Yang}}, \bibinfo {author} {\bibfnamefont {H.~J.}\ \bibnamefont
  {Gao}}, \bibinfo {author} {\bibfnamefont {R.}~\bibnamefont {Zhou}},\ and\
  \bibinfo {author} {\bibfnamefont {G.-q.}\ \bibnamefont {Zheng}},\ }\bibfield
  {title} {\bibinfo {title} {Possible star-of-david pattern charge density wave
  with additional modulation in the kagome superconductor {CsV$_3$Sb$_5$}},\
  }\href {https://doi.org/10.1038/s41535-022-00437-7} {\bibfield  {journal}
  {\bibinfo  {journal} {npj Quantum Materials}\ }\textbf {\bibinfo {volume}
  {7}},\ \bibinfo {pages} {30} (\bibinfo {year} {2022})}\BibitemShut {NoStop}%
\bibitem [{\citenamefont {Song}\ \emph {et~al.}(2003)\citenamefont {Song},
  \citenamefont {Park}, \citenamefont {Koo}, \citenamefont {Lee}, \citenamefont
  {Rhee}, \citenamefont {Bud'ko}, \citenamefont {Canfield}, \citenamefont
  {Harmon},\ and\ \citenamefont {Goldman}}]{PhysRevB.68.035113}%
  \BibitemOpen
  \bibfield  {author} {\bibinfo {author} {\bibfnamefont {C.}~\bibnamefont
  {Song}}, \bibinfo {author} {\bibfnamefont {J.}~\bibnamefont {Park}}, \bibinfo
  {author} {\bibfnamefont {J.}~\bibnamefont {Koo}}, \bibinfo {author}
  {\bibfnamefont {K.-B.}\ \bibnamefont {Lee}}, \bibinfo {author} {\bibfnamefont
  {J.~Y.}\ \bibnamefont {Rhee}}, \bibinfo {author} {\bibfnamefont {S.~L.}\
  \bibnamefont {Bud'ko}}, \bibinfo {author} {\bibfnamefont {P.~C.}\
  \bibnamefont {Canfield}}, \bibinfo {author} {\bibfnamefont {B.~N.}\
  \bibnamefont {Harmon}},\ and\ \bibinfo {author} {\bibfnamefont {A.~I.}\
  \bibnamefont {Goldman}},\ }\bibfield  {title} {\bibinfo {title}
  {Charge-density-wave orderings in {LaAgSb}$_{2}$: An x-ray scattering
  study},\ }\href {https://doi.org/10.1103/PhysRevB.68.035113} {\bibfield
  {journal} {\bibinfo  {journal} {Phys. Rev. B}\ }\textbf {\bibinfo {volume}
  {68}},\ \bibinfo {pages} {035113} (\bibinfo {year} {2003})}\BibitemShut
  {NoStop}%
\bibitem [{\citenamefont {Imre}\ \emph {et~al.}(2007)\citenamefont {Imre},
  \citenamefont {Hellmann}, \citenamefont {Wenski}, \citenamefont {Graf},
  \citenamefont {Johrendt},\ and\ \citenamefont {Mewis}}]{zaac.200700302}%
  \BibitemOpen
  \bibfield  {author} {\bibinfo {author} {\bibfnamefont {A.}~\bibnamefont
  {Imre}}, \bibinfo {author} {\bibfnamefont {A.}~\bibnamefont {Hellmann}},
  \bibinfo {author} {\bibfnamefont {G.}~\bibnamefont {Wenski}}, \bibinfo
  {author} {\bibfnamefont {J.}~\bibnamefont {Graf}}, \bibinfo {author}
  {\bibfnamefont {D.}~\bibnamefont {Johrendt}},\ and\ \bibinfo {author}
  {\bibfnamefont {A.}~\bibnamefont {Mewis}},\ }\bibfield  {title} {\bibinfo
  {title} {Inkommensurabel modulierte {K}ristallstrukturen und
  {P}hasenumwandlungen – {D}ie {V}erbindungen {SrPt$_2$As$_2$} und
  {EuPt$_2$As$_2$}},\ }\href
  {https://doi.org/https://doi.org/10.1002/zaac.200700302} {\bibfield
  {journal} {\bibinfo  {journal} {Z. Anorg. Allg. Chem.}\ }\textbf {\bibinfo
  {volume} {633}},\ \bibinfo {pages} {2037} (\bibinfo {year}
  {2007})}\BibitemShut {NoStop}%
\bibitem [{\citenamefont {Fang}\ \emph {et~al.}(2012)\citenamefont {Fang},
  \citenamefont {Dong}, \citenamefont {Wang}, \citenamefont {Chen},
  \citenamefont {Cheng}, \citenamefont {Shi}, \citenamefont {Zheng},
  \citenamefont {Xu}, \citenamefont {Wang}, \citenamefont {Li},\ and\
  \citenamefont {Wang}}]{PhysRevB.85.184520}%
  \BibitemOpen
  \bibfield  {author} {\bibinfo {author} {\bibfnamefont {A.~F.}\ \bibnamefont
  {Fang}}, \bibinfo {author} {\bibfnamefont {T.}~\bibnamefont {Dong}}, \bibinfo
  {author} {\bibfnamefont {H.~P.}\ \bibnamefont {Wang}}, \bibinfo {author}
  {\bibfnamefont {Z.~G.}\ \bibnamefont {Chen}}, \bibinfo {author}
  {\bibfnamefont {B.}~\bibnamefont {Cheng}}, \bibinfo {author} {\bibfnamefont
  {Y.~G.}\ \bibnamefont {Shi}}, \bibinfo {author} {\bibfnamefont
  {P.}~\bibnamefont {Zheng}}, \bibinfo {author} {\bibfnamefont
  {G.}~\bibnamefont {Xu}}, \bibinfo {author} {\bibfnamefont {L.}~\bibnamefont
  {Wang}}, \bibinfo {author} {\bibfnamefont {J.~Q.}\ \bibnamefont {Li}},\ and\
  \bibinfo {author} {\bibfnamefont {N.~L.}\ \bibnamefont {Wang}},\ }\bibfield
  {title} {\bibinfo {title} {Single-crystal growth and optical conductivity of
  {SrPt${}_{2}$As${}_{2}$} superconductors},\ }\href
  {https://doi.org/10.1103/PhysRevB.85.184520} {\bibfield  {journal} {\bibinfo
  {journal} {Phys. Rev. B}\ }\textbf {\bibinfo {volume} {85}},\ \bibinfo
  {pages} {184520} (\bibinfo {year} {2012})}\BibitemShut {NoStop}%
\bibitem [{\citenamefont {Wang}\ \emph {et~al.}(2014)\citenamefont {Wang},
  \citenamefont {Wang}, \citenamefont {Shi}, \citenamefont {Chen},
  \citenamefont {Chiang}, \citenamefont {Tian}, \citenamefont {Yang},
  \citenamefont {Fang}, \citenamefont {Wang},\ and\ \citenamefont
  {Li}}]{ChinPhysB.23.086103}%
  \BibitemOpen
  \bibfield  {author} {\bibinfo {author} {\bibfnamefont {L.}~\bibnamefont
  {Wang}}, \bibinfo {author} {\bibfnamefont {Z.}~\bibnamefont {Wang}}, \bibinfo
  {author} {\bibfnamefont {H.-L.}\ \bibnamefont {Shi}}, \bibinfo {author}
  {\bibfnamefont {Z.}~\bibnamefont {Chen}}, \bibinfo {author} {\bibfnamefont
  {F.-K.}\ \bibnamefont {Chiang}}, \bibinfo {author} {\bibfnamefont {H.-F.}\
  \bibnamefont {Tian}}, \bibinfo {author} {\bibfnamefont {H.-X.}\ \bibnamefont
  {Yang}}, \bibinfo {author} {\bibfnamefont {A.-F.}\ \bibnamefont {Fang}},
  \bibinfo {author} {\bibfnamefont {N.-L.}\ \bibnamefont {Wang}},\ and\
  \bibinfo {author} {\bibfnamefont {J.-Q.}\ \bibnamefont {Li}},\ }\bibfield
  {title} {\bibinfo {title} {Two-coupled structural modulations in
  charge-density-wave state of {SrPt$_2$As$_2$} superconductor},\ }\href
  {https://doi.org/10.1088/1674-1056/23/8/086103} {\bibfield  {journal}
  {\bibinfo  {journal} {Chin. Phys. B}\ }\textbf {\bibinfo {volume} {23}},\
  \bibinfo {pages} {086103} (\bibinfo {year} {2014})}\BibitemShut {NoStop}%
\bibitem [{\citenamefont {Tsubouchi}\ \emph {et~al.}(2020)\citenamefont
  {Tsubouchi}, \citenamefont {Motoyama}, \citenamefont {Gouchi}, \citenamefont
  {Miyoshi}, \citenamefont {Nishigori}, \citenamefont {Fujiwara}, \citenamefont
  {Mutou},\ and\ \citenamefont {Uwatoko}}]{JPSCP.30.011102}%
  \BibitemOpen
  \bibfield  {author} {\bibinfo {author} {\bibfnamefont {M.}~\bibnamefont
  {Tsubouchi}}, \bibinfo {author} {\bibfnamefont {G.}~\bibnamefont {Motoyama}},
  \bibinfo {author} {\bibfnamefont {J.}~\bibnamefont {Gouchi}}, \bibinfo
  {author} {\bibfnamefont {K.}~\bibnamefont {Miyoshi}}, \bibinfo {author}
  {\bibfnamefont {S.}~\bibnamefont {Nishigori}}, \bibinfo {author}
  {\bibfnamefont {K.}~\bibnamefont {Fujiwara}}, \bibinfo {author}
  {\bibfnamefont {T.}~\bibnamefont {Mutou}},\ and\ \bibinfo {author}
  {\bibfnamefont {Y.}~\bibnamefont {Uwatoko}},\ }\bibfield  {title} {\bibinfo
  {title} {Temperature-pressure magnetic phase diagram of {Ce$_3$TiBi$_5$}},\
  }\bibfield  {booktitle} {\emph {\bibinfo {booktitle} {Proceedings of the
  International Conference on Strongly Correlated Electron Systems
  (SCES2019)}},\ }\href {https://doi.org/10.7566/JPSCP.30.011102} {\bibfield
  {journal} {\bibinfo  {journal} {JPS Conf. Proc.}\ }\textbf {\bibinfo {volume}
  {30}},\ \bibinfo {pages} {011102} (\bibinfo {year} {2020})}\BibitemShut
  {NoStop}%
\bibitem [{\citenamefont {Motoyama}\ \emph {et~al.}(2020)\citenamefont
  {Motoyama}, \citenamefont {Shinozaki}, \citenamefont {Tsubouchi},
  \citenamefont {Kuninaka}, \citenamefont {Nishigori}, \citenamefont {Miyoshi},
  \citenamefont {Fujiwara}, \citenamefont {Gouchi},\ and\ \citenamefont
  {Uwatoko}}]{JPSCP.30.011180}%
  \BibitemOpen
  \bibfield  {author} {\bibinfo {author} {\bibfnamefont {G.}~\bibnamefont
  {Motoyama}}, \bibinfo {author} {\bibfnamefont {M.}~\bibnamefont {Shinozaki}},
  \bibinfo {author} {\bibfnamefont {M.}~\bibnamefont {Tsubouchi}}, \bibinfo
  {author} {\bibfnamefont {M.}~\bibnamefont {Kuninaka}}, \bibinfo {author}
  {\bibfnamefont {S.}~\bibnamefont {Nishigori}}, \bibinfo {author}
  {\bibfnamefont {K.}~\bibnamefont {Miyoshi}}, \bibinfo {author} {\bibfnamefont
  {K.}~\bibnamefont {Fujiwara}}, \bibinfo {author} {\bibfnamefont
  {J.}~\bibnamefont {Gouchi}},\ and\ \bibinfo {author} {\bibfnamefont
  {Y.}~\bibnamefont {Uwatoko}},\ }\bibfield  {title} {\bibinfo {title}
  {Magnetic properties of new antiferromagnetic compound of {Ce$_3$ZrBi$_5$}},\
  }\bibfield  {booktitle} {\emph {\bibinfo {booktitle} {Proceedings of the
  International Conference on Strongly Correlated Electron Systems
  (SCES2019)}},\ }\href {https://doi.org/10.7566/JPSCP.30.011180} {\bibfield
  {journal} {\bibinfo  {journal} {JPS Conf. Proc.}\ }\textbf {\bibinfo {volume}
  {30}},\ \bibinfo {pages} {011180} (\bibinfo {year} {2020})}\BibitemShut
  {NoStop}%
\bibitem [{\citenamefont {Shinozaki}\ \emph {et~al.}(2022)\citenamefont
  {Shinozaki}, \citenamefont {Motoyama}, \citenamefont {Nishigori},
  \citenamefont {Yamaguchi}, \citenamefont {Yamane}, \citenamefont {Mutou},
  \citenamefont {Fujiwara}, \citenamefont {Manago}, \citenamefont {Miyoshi},\
  and\ \citenamefont {Sumiyama}}]{J.Phys.2164.012040}%
  \BibitemOpen
  \bibfield  {author} {\bibinfo {author} {\bibfnamefont {M.}~\bibnamefont
  {Shinozaki}}, \bibinfo {author} {\bibfnamefont {G.}~\bibnamefont {Motoyama}},
  \bibinfo {author} {\bibfnamefont {S.}~\bibnamefont {Nishigori}}, \bibinfo
  {author} {\bibfnamefont {A.}~\bibnamefont {Yamaguchi}}, \bibinfo {author}
  {\bibfnamefont {Y.}~\bibnamefont {Yamane}}, \bibinfo {author} {\bibfnamefont
  {T.}~\bibnamefont {Mutou}}, \bibinfo {author} {\bibfnamefont
  {K.}~\bibnamefont {Fujiwara}}, \bibinfo {author} {\bibfnamefont
  {M.}~\bibnamefont {Manago}}, \bibinfo {author} {\bibfnamefont
  {K.}~\bibnamefont {Miyoshi}},\ and\ \bibinfo {author} {\bibfnamefont
  {A.}~\bibnamefont {Sumiyama}},\ }\bibfield  {title} {\bibinfo {title}
  {Electrical resistivity measurements of antiferromagnetic compound
  {Ce$_3$TiSb$_5$} under pressure},\ }\href
  {https://doi.org/10.1088/1742-6596/2164/1/012040} {\bibfield  {journal}
  {\bibinfo  {journal} {J. Phys.: Conf. Ser.}\ }\textbf {\bibinfo {volume}
  {2164}},\ \bibinfo {pages} {012040} (\bibinfo {year} {2022})}\BibitemShut
  {NoStop}%
\bibitem [{\citenamefont {Nakagawa}\ \emph {et~al.}(2023)\citenamefont
  {Nakagawa}, \citenamefont {Shinozaki}, \citenamefont {Motoyama},
  \citenamefont {Nishigori}, \citenamefont {Fujiwara}, \citenamefont {Manago},\
  and\ \citenamefont {Miyoshi}}]{JPSCP.38.011083}%
  \BibitemOpen
  \bibfield  {author} {\bibinfo {author} {\bibfnamefont {K.}~\bibnamefont
  {Nakagawa}}, \bibinfo {author} {\bibfnamefont {M.}~\bibnamefont {Shinozaki}},
  \bibinfo {author} {\bibfnamefont {G.}~\bibnamefont {Motoyama}}, \bibinfo
  {author} {\bibfnamefont {S.}~\bibnamefont {Nishigori}}, \bibinfo {author}
  {\bibfnamefont {K.}~\bibnamefont {Fujiwara}}, \bibinfo {author}
  {\bibfnamefont {M.}~\bibnamefont {Manago}},\ and\ \bibinfo {author}
  {\bibfnamefont {K.}~\bibnamefont {Miyoshi}},\ }\bibfield  {title} {\bibinfo
  {title} {Single crystal growth of {Ce$_3$ZrSb$_5$} and characterization of
  the physical properties},\ }\bibfield  {booktitle} {\emph {\bibinfo
  {booktitle} {Proceedings of the 29th International Conference on Low
  Temperature Physics (LT29)}},\ }\href
  {https://doi.org/10.7566/JPSCP.38.011083} {\bibfield  {journal} {\bibinfo
  {journal} {JPS Conf. Proc.}\ }\textbf {\bibinfo {volume} {38}},\ \bibinfo
  {pages} {011083} (\bibinfo {year} {2023})}\BibitemShut {NoStop}%
\end{thebibliography}%
\end{document}